\documentstyle[preprint,eqsecnum,aps,epsfig]{revtex}

\tighten
\begin{document}
\newcommand{\Bphi}{\bbox{\phi}}
\newcommand{\BPhi}{\bbox{\Phi}}
\draft
\preprint{DESY 96-227}
\title{A Monte Carlo Renormalization Group Approach \\ 
       to the Bak-Sneppen model} 
\author{Bernhard Mikeska}
\address{II. Institut f\"ur Theoretische Physik, Universit\"at Hamburg, \\
Luruper Chaussee 149, D-22761 Hamburg}
\date{October 31, 1996}
\maketitle
\begin{abstract}
A recent renormalization group approach to a modified Bak-Sneppen model is 
discussed. We propose a self-consistency condition
for the blocking scheme to be
essential for a successful RG-method applied to 
self-organized criticality. A new method realizing the RG-approach to the 
Bak-Sneppen model is presented. It is based on the Monte-Carlo importance 
sampling idea. The new technique performs much faster than the original 
proposal. Using this technique we cross-check and improve previous results.
\end{abstract}

\newpage

%\narrowtext
\section{Introduction}
Biological evolution has been a candidate for 
self-organized criticality (SOC) \cite{btw87} for a long time.
The Bak-Sneppen (BS) model \cite{bs93} was the first model describing 
biological evolution as a self-organized critical phenomenon.
A new renormalization group (RG) approach to calculate 
critical exponents of the BS-model has been presented first in \cite{mar94b}.
This note presents a Monte Carlo technique to realize the approach 
of \cite{mar94b} and provides thereby an easy way of cross-checking
the novel RG-method. It accelerates and simplifies the calculation by using a 
Monte-Carlo inspired technique. We are able to improve
the results given in \cite{mar94b}.
The paper is organized as follows: in section \ref{sec:RG} we redefine the
BS-model and review the main ideas of the RG-approach. The main tool 
of this approach is the Run-Time-Statistics (RTS) technique, first presented 
in \cite{mar94a}. We discuss it in section \ref{sec:RTS}.
Section \ref{sec:MCRG} introduces our new technique to perform the RG-approach
more efficiently. The fixed point properties are given and basic critical 
exponents are calculated.

\section{The RG-approach to the BS-model}
\label{sec:RG}

Following Ref.\ \cite{mar94b} we modify the BS-model and define a 
{\em left-or-right} L/R-BS-model:
Consider $N$ real variables $\Bphi^{(t)} = (\phi_i^{(t)}),\; i \in \{1,\ldots,N\}$ with
values $0.0 \le \phi_i^{(t)} < 1.0$. At each time $t$ determine index $i(t)$
indicating the smallest value $\phi_{i(t)}$ among all $\phi_j$. Replace the 
value $\phi_{i(t)}$ by a new value $x$ with probability $a(x)$, $\int a(x) dx =
1$. Choose with equal probability the left or right neighbor of the active site
$i(t)$ and replace it by a new value $y$ with probability $b(y)$, 
$\int b(y) dy = 1$ \cite{remark3}. 
After equilibration almost all variables $\phi_j$ have a higher value than 
some $\rho_c$.
A $\rho$-avalanche is defined to start at time $t_i$ with the 
minimal value $\phi_{i(t_i)} = \rho$, lasts as long as 
consecutive minimal values are smaller than $\rho$ and stops at $t_f$ if
$\phi_{i(t_f)} \ge \rho$. The temporal extension is given by $s\equiv t_f-t_i$.
The spatial extension $\xi$ is defined as the largest extent of active sites 
involved, $\xi \equiv \max_{t_i \le t_1,t_2 \le t_f}
|i(t_1) - i(t_2)|$. Avalanches keep themselves running by generating
variables smaller than $\rho$. If this number of variables smaller than
$\rho$,
$n_t(\rho) \equiv \sum_i \Theta(\rho -\phi_i^{(t)})$, decreases with time 
during a 
$\rho$-avalanche the avalanche is called sub-critical and will die out. On the
other hand, if $n_t(\rho)$ increases, the avalanche is called super-critical 
and will last forever. If $n_t(\rho)$ stays constant we observe a critical 
avalanche. Critical avalanches obey power laws for their spatial and
temporal size, $P_\xi(\xi) \sim \xi^{-\tau_\xi}$ and $P(s) \sim s^{-\tau}$ 
respectively. The connection between spatial and temporal size is given by 
$s\sim\xi^z$. The critical exponents $\tau_\xi$, $\tau$ and $z$ are connected 
via the scaling relation $z\tau = \tau_\xi - 1 + z$. 
Extensive numerical studies have been performed to calculate these exponents
for the original BS-model\cite{gra95,pmb95,mik96a}. 

The goal of a RG-approach is threefold: calculate critical exponents in
an independent analytical or semi-analytical way; prove the 
attractive nature of the critical state; demonstrate the concept of 
universality. 
The RG-approach \cite{mar94b} considers small 
avalanches as objects to be integrated out and provides a mapping of larger
avalanches onto smaller ones. This goes in parallel with a modification of the
dynamical rules at coarser scales.

%\enlargethispage{-2\baselineskip}

\begin{figure}
\center\epsfig{file=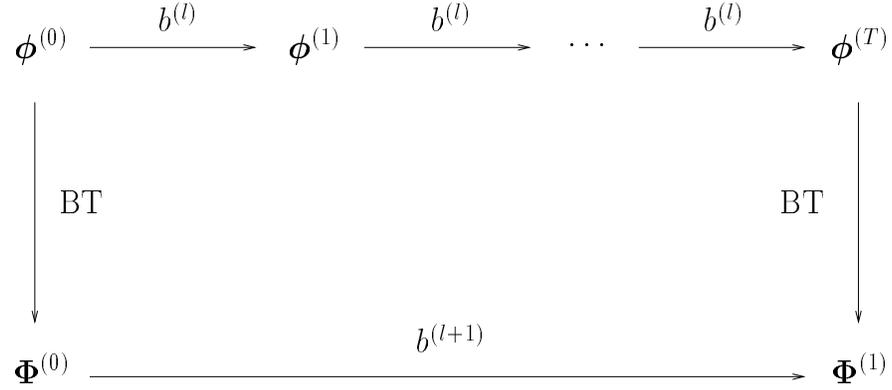,width=15.5cm,bbllx=0,bblly=400,bburx=596,
               bbury=842}
\caption{Concept of blocking in space and time using a block transformation BT.
$T$ updates on
the fine level $l$ are performed until four neighboring variables are changed.
The result of this process defines a contribution to one time step on the
coarse level $l+1$.}
\label{fig:rgComDia}
\end{figure}

To be more precise, let us denote the fine scale with index $(l)$, the next 
coarser scale with $(l+1)$. Block variables are $\Phi_I$, fine variables are 
$\phi_i$. Using a block factor of two, the simplest block transformation 
conserving the spatial avalanche structure is
\begin{equation}
\label{eq:rgBlockDef}
\Phi_I^{(t')} = \min \{\phi_{2i}^{(t)}, \phi_{2i+1}^{(t)}\} \; .
\end{equation}
Then the dynamics on the coarser level is again based on the selection of the
minimum. To perform one update on the coarse grid, i.e. updating $\Phi_{I(t)}$
and either the left or right neighbor $I(t) \pm 1$, we have to consider a 
process with four neighboring variables on the finer scale being updated. 
Thus a dynamical blocking of time occurs and we have different 
time scales on the fine and coarse level. 
The mapping connecting both time scales is dynamic, i.e. it depends on the 
stochasticity of the temporal evolution. 
This will provide us with a modified dynamics based on new
distributions $a^{(l+1)}(x)$ and $b^{(l+1)}(x)$. 
The initial and final state of one updating step on the 
coarse level is obtained by applying equation (\ref{eq:rgBlockDef}) 
to the initial and final configuration of the corresponding fine level process.
The spatial and temporal blocking of the dynamical process is schematically 
shown in Fig.\ \ref{fig:rgComDia}. The block transformation and 
the modified dynamics have to be designed such that it makes statistically
no difference for the final state $\Phi^{(1)}$ if we apply first $T$ 
fine updates and then the block transformation or if we apply first the 
block transformation and then perform one coarse update using the coarse 
dynamics: the blocking diagram has to 
be commutative for a successful RG-approach. 

Assume that we already know the 
probability distribution $\tilde{a}^{(l+1)}(x)$ and $\tilde{b}^{(l+1)}(x)$ 
of variables $\Phi_{I(t)}$ and $\Phi_{I(t)\pm 1}$
in the final state on the coarse level.
It is then possible \cite{mar94b} to perform a variable transformation 
$x \rightarrow y(x)$ 
using $y(x) = \int_0^x \tilde{a}^{(l+1)}(x') dx'$ 
leading to $a^{(l+1)}(y) = \Theta(y) \Theta(1-y)$ and 
$b^{(l+1)}(y) = \left . \frac{\tilde{b}^{(l+1)}(x)}
                {\tilde{a}^{(l+1)}(x)} \right |_{y(x)=y}$ .
This reduces the possible RG-proliferation 
to the distribution $b(x)$. All other rules (search for the minimum,
choose left or right neighbor, replace the value of the minimal variable with
a uniformly distributed one) are invariant under the RG-transformation. 

Scale invariant behavior is expected only for avalanches that are critical, 
i.e. $\phi_{i(t_i)}^{(l)} = \rho_c^{(l)}$. Therefore one has to consider only
critical processes for the mapping of a sequence of fine updates onto one 
coarse update. Only processes $\alpha$ with four neighboring sites
being updated after $T_\alpha$ time steps are relevant. Processes that stop 
earlier because $\phi_{i(t')}^{(l)} \ge \rho_c^{(l)}$ for $t' > t_i$ do not 
contribute since they represent small fluctuations with avalanche size 
$S \le 3$ and are neglected in the RG step. The parameter $\rho_c^{(l)}$
is determined from the condition that $n_t(\rho_c^{(l)},b^{(l)})$
should stay constant for critical avalanches, see eq.\ (\ref{eq:solRho})
below.

\section{Realization of the RG-approach using Run-Time-Statistics}
\label{sec:RTS}

We now want to discuss the Run-Time-Statistics approach developed 
in \cite{mar94a}.
Denote the probability distribution of $\phi_i$ at time $t$ by
$p_{i,t}(x)$. The conditional probability that the minimal variable is located 
at site $i$ is then
\begin{equation}
\label{eq:rtsMin}
p_{\text{min},i}^t(x) = p_{i,t}(x) \prod_{j \neq i} \int_x^1 p_{j,t}(y)dy \;\; .
\end{equation}
For the blocking procedure we are interested in the probability
that $\phi_{i}^{(t)}$ is the minimal site {\it and}
that the avalanche does not stop, i.e. $\phi_{i}^{(t)} < \rho$, which is 
\begin{equation}
\label{eq:rtsMinProb} 
\mu_{i,t}(\rho) = \int_0^\rho p_{\text{min},i}^t(x) dx \;\; .
\end{equation} 
Exploiting the information about the position of the minimal value $\phi_i$,
$\phi_i < \rho$, the probability distribution $p_{j,t}(x)$ of all other sites 
$j$ modifies in the next time step to
\begin{eqnarray}
\label{eq:rtsNotMin}
  p_{j,t+1}(x) & = & {\cal N}p_{j,t}(x) \int_0^\rho p_{i,t}(y) \Theta(x-y)
                                                      \nonumber \\ 
  & & \times \prod_{k \neq i,j} \int_y^1 p_{k,t}(z) dz dy \;\; .
\end{eqnarray} 
$\cal{N}$ is a normalization factor.
The probability distribution of the minimal site $i(t)$ and its left or
right neighbor change according to the rules of the L/R-BS-model to 
\begin{mathletters}
\label{eq:rtsIteration}
\begin{eqnarray}
p_{i(t),t+1}(x) &=& \Theta(x)\Theta(1-x)\\ 
p_{i(t)\pm 1,t+1}(x) &=& b(x) \; .
\end{eqnarray}
\end{mathletters}
Note that equation (\ref{eq:rtsNotMin}) is different \cite{remark1} 
from the corresponding equations in \cite{mar94b,mar94a}. 
This leads to different results of the fixed point properties, see below.  
Using the iterative rules (\ref{eq:rtsNotMin}) and (\ref{eq:rtsIteration})
we can calculate the weight $\nu_\alpha$ of a given process $\alpha$ 
contributing to the blocking procedure
\begin{equation}
\label{eq:rtsProcessWeight}
\nu_\alpha(\rho) = \frac{1}{2^{T_\alpha}}\prod_{t=1}^{T_\alpha} \mu_{i_\alpha(t),t}(\rho) \;\; .
\end{equation}
The final probability distributions in the
four updated variables contribute to the
blocked distributions $\tilde{a}$ and $\tilde{b}$ at the next coarser level
according to the weight of process $\alpha$.
Using the variable transformation described in \cite{mar94b}, $\tilde{a}(x)$ may be 
rescaled to a uniform distribution in the interval $[0,1]$ and the next
renormalization procedure can be iterated. 

\section{The Monte-Carlo realization of the RG-approach}
\label{sec:MCRG}

To realize the renormalization group approach using the RTS-technique, 
one has to calculate the weights and final probability distributions of 
all contributing processes for each iteration during the computation of 
$\rho_c$, using equation (\ref{eq:solRho}) below. Then one has to perform 
the same task for all paths using the value $\rho_c$. Finally the  
block transformation (\ref{eq:rgBlockDef}) has to be applied. 
However, there is a huge amount of contributing processes: 
up to a length of $T_\alpha \le 20$ we estimated a total number of $O(10^{12})$
contributing processes. It is also clear that most of them, especially of the 
longer processes (e.g. the ones with the minimal site at the same place for a 
long time) are of very small probability. This observation reminds to the 
problems in ''simple sampling'' Monte Carlo algorithms, calculating properties
and probabilities of all possible states in phase space. It is more convenient
to use an ''importance sampling'' method generating the contributing processes 
$\alpha$ according to their weight $\nu_\alpha$. 
Then all generated processes of the ensemble 
contribute with equal probability. 
In other words, we evaluate RTS-integrals using the Monte-Carlo 
importance sampling method. 

To generate a single relevant process $\alpha$ using distribution
$b^{(l)}(x)$, we start with a variable $\phi_{i_\alpha(0)} = \rho$,
and apply the rules of the L/R-BS-model until
four neighboring sites are updated. For this process we count
the number of variables smaller than $\rho$ after the first time
step as well as in the final state
\begin{mathletters}
\begin{eqnarray}
  n_1^{(\alpha)} &= \sum_i \Theta(\rho - \phi_{i_\alpha}^{(t=1)}) \\
  n_{T_\alpha}^{(\alpha)} &= \sum_i \Theta(\rho - \phi_{i_\alpha}^{(t=T_\alpha)}) \; .
\end{eqnarray}
\end{mathletters}
If $n_1^{(\alpha)} = 0$ or $n_{T_\alpha}^{(\alpha)} = 0$, 
the generated process is 
not a relevant process, since it stopped, i.e. $\phi_{i(t)} > \rho$, 
before four neighboring sites are updated. Using this notation we can easily 
write down $n_{t=1}(\rho,b^{(l)})$ and $n_{t=t_{\text{final}}}(\rho,b^{(l)})$ 
obtained from the generation of an ensemble of $N$ processes \cite{remark2}
\begin{mathletters}
\label{eq:nDef}
\begin{eqnarray}
n_{t=1}(\rho,b^{(l)}) &=& \frac{1}{N} \sum_\alpha n_1^{(\alpha)} \\
n_{t=t_{\text{final}}}(\rho,b^{(l)}) &=& \frac{1}{N} \sum_\alpha n_{T_\alpha}^{(\alpha)} 
\end{eqnarray}
\end{mathletters}
Since we are interested only in critical avalanches we have
$n_t(\rho_c^{(l)},b^{(l)}) = \text{\it const.}$ and obtain $\rho_c^{(l)}$ as solution of 
\begin{equation}
\label{eq:solRho}
n_1(\rho_c^{(l)},b^{(l)}) = n_{t_{\text{final}}}(\rho_c^{(l)},b^{(l)})\;\; .
\end{equation}
Using the secant method it converges in about five iterations.
Once we know the critical $\rho_c^{(l)}$ at level $l$ for given
$b^{(l)}$, we are able to evaluate all {\it critical} relevant processes 
$\alpha$ starting with $\phi_{i_{\alpha}(0)} = \rho_c^{(l)}$ for their 
contribution to the blocked distributions $\tilde{a}^{(l+1)}(x)$ and 
$\tilde{b}^{(l+1)}(x)$. A relevant process $\alpha$
contributes with
\begin{mathletters}
\begin{eqnarray}
\tilde{a}_{\alpha}^{(l+1)}(x) &=& \delta(x - m_{\alpha}),\;
  m_{\alpha} = \min\{\phi_{2j}^{(t_\alpha)},\phi_{2j+1}^{(t_\alpha)}\} \\
\tilde{b}_{\alpha}^{(l+1)}(x) &=& \delta(x - n_{\alpha}),\;
  n_{\alpha} = \min\{\phi_{2k}^{(t_\alpha)},\phi_{2k+1}^{(t_\alpha)}\} \; .
\end{eqnarray}
\end{mathletters}
The first active site $i_{\alpha(t=0)}$ has index $2j$ or $2j+1$ 
and the pair $(2k,2k+1)$ is the right or left neighbor of $(2j,2j+1)$. 
The four indices $2j,2j+1,2k,2k+1$ have been updated in process $\alpha$.
For an ensemble of $N'$ relevant processes $\alpha$, each occurring with a
probability $\nu_\alpha(\rho_c^{(l)})$, we obtain 
\begin{mathletters}
\begin{eqnarray}
\tilde{a}^{(l+1)}(x) &=& 
      \frac{1}{N'}\sum_\alpha\tilde{a}_{\alpha}^{(l+1)}(x) \\
\tilde{b}^{(l+1)}(x) &=& 
      \frac{1}{N'}\sum_\alpha\tilde{b}_{\alpha}^{(l+1)}(x) \;\; . 
\end{eqnarray}
\end{mathletters}
To these preliminary distributions we apply the variable transformation
described above, thereby switching back to a uniform distribution 
$a^{(l+1)}(x)$ and a transformed distribution $b^{(l+1)}(x)$. Now the 
described procedure may be iterated. 

An ensemble of processes $\alpha$, each occuring with probability 
$\nu_\alpha$, is generated by applying the rules of the L/R-BS-model 
to processes starting with $\phi_{i_{\alpha}(0)} = \rho$ 
as long as not four neighboring sites have been updated. Once this happens, 
we stop the process and save its initial and final state. We reject 
non-relevant processes. 
Then the process $\alpha$ 
occurs automatically with its correct weight $\nu_\alpha$ without the need 
of an explicit calculation of $\nu_\alpha$ using RTS. 
Then we start the same procedure again for the next process that is
generated completely independently from the previous one.  

Being considerably simpler our method is able to reproduce the results 
of \cite{mar94b}
with good accuracy 
in about one hour CPU-time on a workstation. Moreover, there is no need to 
extrapolate the results from $T_{\text{max}} = 20$ to $T_{\text{max}} = 
\infty$ as done in \cite{mar94b}. 
In our approach we effectively consider $T_{\text{max}} = \infty$, 
since we allow arbitrary long relevant processes.

On the other hand, since we 
generate a finite number $N$ of processes, we observe statistical
fluctuations in the resulting distribution $b^{(l)}(x)$, whereas
the method \cite{mar94b} gives in this sense an exact result (neglecting small 
errors due to the numerical integration routines). But this
is a relatively small drawback compared to the advantage of speeding
up the method, thereby having an effective method at hand to cross-check 
previous results.

\begin{figure}
\center\epsfig{file=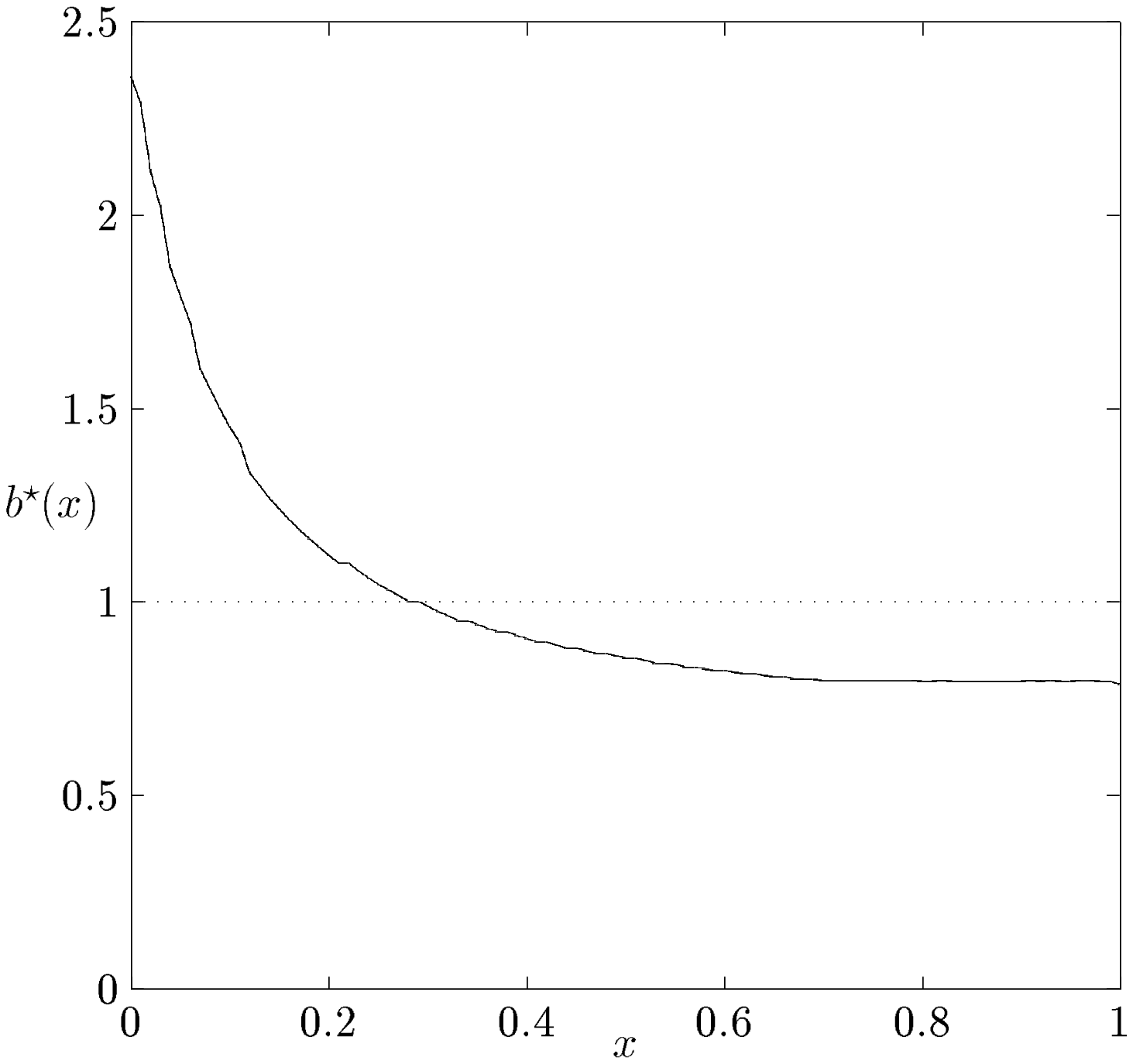,width=15.5cm,bbllx=0,bblly=200,bburx=596,
               bbury=842}
\caption{The fixed point  distribution $b^\star(x)$ is observed 
already after one renormalization step. For each RG-step $10^9$ relevant 
processes have been generated.}
\label{fig:mcrgResultFP}
\end{figure}

Starting with a distribution $b^{(0)}(x) = \Theta(x) \Theta(1-x)$
at the finest level $l=0$, we observe that the distribution $b^{(l)}(x)$
converges very fast. It reaches its fixed point shape already after one 
renormalization step, see Fig.\ \ref{fig:mcrgResultFP}. 
Correspondingly the value of $\rho_c^{(l)}$ converges also very fast to its 
fixed point value $\rho_c^\star = 0.5954$.
For the calculation of $b^{(l)}(x)$
we evaluated $10^9$ relevant processes at each renormalization step. For
the determination of $\rho_c^{(l)}$ we evaluated $10^7$ processes per 
iteration of the secant method and renormalization step.
The results of \cite{mar94b} differ significantly which is
probably due to the error in one of the RTS-iteration equations \cite{remark1}.

The probability distribution $P^\star(T)$ for the length $T$ of relevant 
processes, corresponding to the fixed point properties $b^\star$ and 
$\rho_c^\star$, decays exponentially: 
$P^\star(T)\sim e^{-\text{const}\times T}$. The minimal length of a 
relevant process is $T=3$. At $T=30$ the probability is of order $10^{-5}$.

To compute the critical exponents from the fixed point properties we
follow the discussion of \cite{mar94b} and use equations (5) and (6) given 
therein. For details the reader is referred to \cite{mar94b}.  
From the probability distribution $P^\star(T)$ we obtain the exponent 
$z = 2.336(1)$. For the calculation of exponent $\tau_\xi$, describing the 
spatial scaling of avalanche sizes, the probability $K^\star$, that the site
with minimal value is always the same (until the avalanche stops), 
has to be known. The value of $K^\star$ may be obtained again using 
our Monte-Carlo approach. We have to count the fraction of processes with
the minimum at the same site at each time step until they stop due to the
condition $\phi_{i_\beta(T_\beta)} \ge \rho_c^\star$. Using the
fixed point distribution $b^\star(x)$,  
we find $K^\star = 0.1827(1)$, from which follows 
$\tau_\xi = 1.2911(1)$.
Using the scaling relation, connecting $\tau_\xi$, $\tau$, and $z$, and our 
result for the dynamical exponent $z$ we obtain for
$\tau = 1.1246(1)$.

The values of exponents obtained via the renormalization group approach
and via numerical simulation are compared in Table \ref{tab:mcrgExp}. 
Due to the probabilistic nature of our approach the fixed point distributions
fluctuates a little. The relative error in $\rho_c^\star$, $z$, $K^\star$,
$\tau_\xi$ and $\tau$ induced by this fluctuation has been estimated of 
order of $10^{-3}$. The results obtained by numerical integration of the 
RTS-equations given in \cite{mar94b} differ significantly due to the error 
in one of the RTS-equations.
The results of the Monte Carlo RG-approach show a coincidence up to 4\% with 
numerical values for the exponents of the original BS-model. It is assumed 
that the L/R-version is in same universality class as the original BS-model.  
    
\section{Conclusions}

We have introduced a new Monte Carlo renormalization group method 
for the L/R-Bak-Sneppen model. Previous results for basic critical exponents 
are improved. We have proposed a self-consistency condition
for the blocking diagram in RG-methods applied to  SOC systems.

\section*{Acknowledgments}

I would like to thank G.\ Mack and Y.\ Xylander for stimulating
discussions. This work was supported in parts by the Deutsche 
Forschungsgemeinschaft and the Studienstiftung des deutschen Volkes.
  
\begin{table}
\caption{\protect\small Comparison of results for critical exponents
using the renormalization group approach and simulations. The
renormalization group approach applies to the L/R-version of the BS-model.
RTS-approach results are given in \protect\cite{mar94b}}.
\begin{tabular}{cccc}
   & \multicolumn{2}{c}{RG method} & Simulation\\ 
   & MC approach & RTS approach & original BS \\ \tableline
   $\tau$    & $1.1246(1)$\tablenotemark[1] & $1.1204$\tablenotemark[1] 
                                              & 1.08(1) \\ 
   $\tau_\xi$ & 1.2911(1) & 1.2766 & \\ 
   $z$       & 2.336(1) & 2.2975 & $2.43(1)$\tablenotemark[2] \\ 
\end{tabular}
\tablenotetext[1]{These results have been obtained from the scaling relation.}
\tablenotetext[2]{This result is given in \protect\cite{pmb95}.}
\label{tab:mcrgExp}    
\end{table}

\newpage

\begin{thebibliography}{99}
\bibitem{remark1} Marsili did not take into account the fact, that 
$m_{i,t}(x)$ and $p_{j,t}(y)$, $j \neq i$, (using the notation of 
\cite{mar94b}) are not mutually independent and it is not allowed to multiply 
them to obtain the new conditional probability.
\end{thebibliography}
\end{document}